\documentclass[12pt]{iopart}

\usepackage{graphicx}
\usepackage{color}
\usepackage{enumerate}

\begin{document}

\title[]{The influence of strong magnetic fields on proto-quark stars}

\author{V.~Dexheimer}
%\email{vantoche@gettysburg.edu}
\address{Depto de F\'{\i}sica - CFM - Universidade Federal de Santa
Catarina  Florian\'opolis\\SC - CP. 476 - CEP 88.040 - 900 - Brazil}
\address{Department of Physics, Kent State University, Kent, OH 44242 United States}

\author{D. P. Menezes}
%\email{debora.p.m@ufsc.br}
\address{Depto de F\'{\i}sica - CFM - Universidade Federal de Santa
Catarina  Florian\'opolis\\SC - CP. 476 - CEP 88.040 - 900 - Brazil}
%\affiliation{Physics Department, UFSC, Florianopolis, Brazil}

\author{M. Strickland}
%\email{mstrickl@gettysburg.edu}
\address{Department of Physics, Kent State University, Kent, OH 44242 United States}

\begin{abstract}
We analyze different stages of
magnetized quark star evolution incorporating baryon number
conservation and using an anisotropic energy momentum tensor. 
The first stages of the evolution are simulated through the
inclusion of trapped neutrinos and fixed entropy per particle, while
in the last stage the star is taken to be deleptonized and cold. We
find that, although strong magnetic fields modify
quark star masses, the evolution of isolated stars needs to be constrained
by fixed baryon number, which necessarily lowers the possible star masses.
Moreover, magnetic field effects, measured
by the difference between the parallel and perpendicular pressures,
are more pronounced in the beginning of the star evolution, when
there is a larger number of charged leptons and up quarks. 
We also show that having a spatially varying magnetic field allows for larger magnetic fields to be supported.
\end{abstract}

\pacs{26.60.Kp, 21.65.Qr, 97.10.Cv, 97.10.Ld}
\maketitle

\section{Introduction}

Neutron stars are compact objects with masses of the order of 1-2
$M_\odot$, radii around 10 km, and a temperature of  approximately
$10^{11}$ K at birth,  which cool rapidly by emitting neutrinos. In
conventional neutron star models, they are comprised of hadronic
matter and leptons. Nevertheless, according to the Bodmer-Witten conjecture, the
true ground state of matter could be deconfined up, down, and strange quarks in environments
where densities are very large
\cite{Bodmer:1971we,Witten:1984rs}.
This hypothesis led to the possibility that neutron stars could
also be strange stars or quark stars, comprised of quarks and
leptons \cite{Itoh:1970uw,PeresMenezes:2005bc}. The true nature of these compact
stars remains a source of speculation and further constraints are
necessary so that models can be ruled out.

Magnetars have extremely high magnetic fields of up to $B\sim10^{15}$ G
on the surface \cite{c1,c2,c3,c4}. Nevertheless, all pulsars have relatively strong magnetic fields \cite{b1,b2}, and for this reason, any complete analysis of pulsar features should include magnetic field effects. Unfortunately, all realistic calculations of magnetic field limits in the center of stars are model dependent to some extent, even when the virial theorem is considered. Some results along this line can be found in Refs.~\cite{Bocquet:1995je,Cardall:2000bs,1991ApJ...383..745L,Chakrabarty:1997ef,Bandyopadhyay:1997kh,Broderick:2001qw,Ferrer:2010wz,Malheiro:2004sb} and have limits ranging between $B\sim10^{17}-10^{20}$ G. At such high magnetic fields, the thermodynamical as well as the hydrodynamical properties of matter become anisotropic \cite{Chaichian:1999gd,PerezMartinez:2005av,PerezMartinez:2007kw,Huang:2009ue,Paulucci:2010uj,Isayev:2011zz,Strickland:2012vu,Sinha:2010fm} and, therefore, need to be carefully handled.

Although much of the information we have about compact stars
  comes from pulsars in binary systems, most pulsars
  are isolated objects, in which case one needs to take care to 
properly enforce baryon number conservation.
This happens because, differently from pulsars in
binary systems which can accrete matter from a companion, isolated
stars cannot and, therefore, conserve baryon number. In this case, as was previously shown in
Refs.~\cite{1995NuPhA.588..365T,Prakash:1996xs,GondekRosinska:2000ie}, there exist  windows of stability which only allow
for stars with particular values of baryon number to exist at any given time. The different snapshots of the evolution are simulated
though different entropies per particle and trapped neutrinos. It was shown in Ref.~\cite{Dexheimer:2008ax} following such an
analysis that the maximum mass that can possibly be attained by
hadronic stars decreases with time.

There have been some real-time calculations of the evolution of proto-neutron stars performed
in Refs.~\cite{Burrows:1986me,Pons:2001ar,pons/pons,marcelo} utilizing Boltzmann transport equations with realistic
neutrino cross sections. Although a very interesting preliminary calculation 
of the neutrino absorption cross section influenced by magnetic field
effects was performed in Ref. \cite{maruyama} for neutron stars, to the best of our knowledge, such
early time simulations have not been performed including magnetic
field effects on the equation of state itself.

Considering the equation of state, there have been previous studies of proto-quark and proto-hybrid stars \cite{Steiner:2001rp,%
Pons:2001ar,Gupta:2002hj,Berdermann:2005yn,Shen:2005vh,Nicotra:2006eg,Burgio:2007eg,%
Yasutake:2009kj,Yasutake:2010ur,Bombaci:2011mx,Lugones:2011xv,Shao:2011nu,Chen:2012zx}.
Others have considered proto-neutron stars subject to a magnetic field \cite{Pal:1998jb,Bednarek:2002hb}
and zero temperature quark stars in a magnetic field using the bag model \cite{Chakrabarty:1996te,Isayev:2012sv}.  
In this paper we perform the first study of the effects of magnetic fields on finite temperature proto-quark stars incorporating the 
effect of baryon number conservation.  As mentioned above, this is important when considering isolated proto-quark
stars.  In addition, we include the pressure anisotropies generated by all relevant degrees of freedom (quarks, leptons, and neutrinos) and the magnetic field.  

Using the MIT bag model \cite{Chodos:1974je}, we determine the effect of strong magnetic fields on three 
snapshots of quark star evolution.
Our task includes a comprehensive investigation of the pressure anisotropy
at zero temperature and at fixed entropy per baryon, including finite temperature effects. In most of the previous
works on neutron stars subject to strong magnetic fields, the anisotropic
pressures were either disregarded \cite{nosso_NJL,nosso_NJL2,luiz,aziz,Rabhi:2009ih} or only considered at zero temperature
\cite{PerezMartinez:2005av,PerezMartinez:2007kw,Huang:2009ue,Paulucci:2010uj}. Here we ignore the effects of the quark anomalous magnetic moment (AMM) and focus on the effect of Landau quantization only. 

The structure of the paper is as follows.  In Sec.~\ref{sect:form} we review the formal setup of the problem
and present expressions for the bulk properties of the system.  In Sec.~\ref{sect:results} we present our results
for the transverse and longitudinal pressures and analyze the effect on the mass-radius relationship and star
properties at three different moments in its evolution.  In Sec.~\ref{sect:conclusions} we present our conclusions
and an outlook for the future.

\section{Formalism}
\label{sect:form}

The Lagrangian density for the bag model including the magnetic field
is given by
\begin{eqnarray}
{\cal{L}}&=&\Big[\bar{\Psi}_q\left(i\gamma^\mu\partial_\mu-{Q_e}_qe\gamma^\mu A_\mu-m_q\right)\Psi_q-{\mathcal B}\Big]\Theta_V\nonumber\\&-&\frac{1}{16\pi}F_{\mu\nu}F^{\mu\nu}+\bar{\Psi}_l\left(i\gamma^\mu\partial_\mu-{Q_e}_le\gamma^\mu A_\mu-m_l\right)\Psi_l \ ,
\label{lag}
\end{eqnarray}
where $\Psi_q$ stands for the three light quark fields, ${Q_e}_i$ is the electric charge of each particle in multiples of the electron charge, $e=0.08$ is the
electron charge (in Gaussian natural units), $A^\mu$ is
the electromagnetic field and $\Theta_V$ is a step function defined as
$1$ inside the bag and $0$ outside. The bag constant value is chosen
to be ${\mathcal B}=(154\ \rm{MeV})^4$ in accordance with an analysis of
stability made for magnetized proto-quark stars
\cite{us}. Note that more sophisticated models include bag ``constants" that are functions of the chemical potential or of the magnetic field itself \cite{Reinhardt:1986tv,Adami:1993tp,Liu:2001em,Burgio:2002sn,Mallick:2012wa}. Other models include 
density dependent quark masses instead of constant bag
parameters \cite{qmdd,qmdd2,qmdd3,us}. In this kind of models, a pressure term that depends on
the baryonic density (or chemical potential) or on the
  magnetic field itself is present.  
We do not follow such prescriptions in order to keep the density dependence of the model as  simple and clear as possible.  $F_{\mu\nu}=\partial_\mu A_\nu-\partial_\nu A_\mu$ is the electromagnetic field tensor. Note that there is a factor $1/{4\pi}$ multiplying $F_{\mu\nu}F^{\mu\nu}$ due to our choice of units. Finally, $\Psi_l$ stands for the electron, muon, and electron and muon neutrino fields.

The magnetic field is taken to be
parallel to the $z$ direction. For details of the calculation of the bulk properties of a Fermi gas we refer to Refs.~\cite{Strickland:2012vu,PerezMartinez:2005av,PerezMartinez:2007kw,%
Paulucci:2010uj,nosso_NJL,nosso_NJL2,luiz,aziz,Rabhi:2009ih,Chiu:1968zz,Canuto:1969ct,Canuto:1969cs,Canuto:1969cn}. The pressure and energy density in the bag model are the pressure and energy density of a non-interacting Fermi gas with the bag constant subtracted and added, respectively.  For the charged particles, the matter contribution to the pressure and energy density for finite and zero temperature in the presence of a constant magnetic field, $B$, in the $z$ direction are 
\begin{eqnarray}
P_{m_\|}=\sum_{i,\nu} \frac{\gamma_i(\nu)}{2 \pi^2}|{Q_e}_i|eB \int \!\! \frac{k_i^2}{\sqrt{k_i^2+\bar{m}_i^2}} ({f_+}_i+{f_-}_i) dk - {\mathcal B} \  ,
\end{eqnarray}\
\begin{eqnarray}
P_{m_\|}(T=0)=\sum_{i,\nu} \frac{\gamma_i(\nu)}{2 \pi^2}|{Q_e}_i|eB \frac 1 2 \left[\bar{k}_{Fi} \mu_i-\bar{m}_i^2\ln{\left|\frac{\bar{k}_{Fi}+\mu_i}{\bar{m}_i}\right|}\right]-{\mathcal B} \ ,
\end{eqnarray}
\begin{eqnarray}
\epsilon_{m}&=&\sum_{i,\nu} \frac{\gamma_i(\nu)}{2 \pi^2}|{Q_e}_i|eB \int \sqrt{k_i^2+\bar{m}_i^2} ({f_+}_i+{f_-}_i) dk + {\mathcal B} \, ,
\end{eqnarray}
\begin{eqnarray}
\epsilon_{m}(T=0)&=&\sum_{i,\nu} \frac{\gamma_i(\nu)}{2 \pi^2}|{Q_e}_i|eB \frac 1 2 \left[\bar{k}_{Fi} \mu_i+\bar{m}_i^2\ln{\left|\frac{\bar{k}_{Fi}+\mu_i}{\bar{m}_i}\right|}\right]+{\mathcal B} \ ,
\end{eqnarray}
where $\gamma_i(\nu)$ is the degeneracy of each particle taking into account spin and/or number of colors $i$ (6 for quarks and 2 for electrons and muons except when $\nu=0$ -- see below), $\bar{k}_{Fi}\equiv\sqrt{\mu_i^2-\bar{m}_i^2}$ the Fermi momentum of each particle modified by the magnetic field, and $\bar{m}_i=\sqrt{m_i^2+2|{Q_e}_i|eB\nu}$ the mass of each particle modified by the magnetic field with $m_i$ being the bare mass of each particle (with $m_{u,d}=5$ MeV and $m_s=150$ MeV). The distribution functions for the particles ($f_+$) and anti-particles ($f_-$) are ${f_\pm}_i=1/[e^{(\sqrt{\bar{k}_i^2+\bar{m}_i^2} \mp \mu_i)/T}+1]$, where $T$ is the temperature.

The chemical potential of each particle $\mu_i$ is calculated from the independent chemical potentials (baryon $\mu_B$, charge $\mu_q$ and lepton $\mu_l$,), each of which is related to a conserved quantity (baryon number, electric charge and lepton fraction)
\begin{eqnarray}
\mu_i={Q_B}_i \mu_B+ {Q_e}_i (\mu_q + \mu_l) + {Q_l}_i \mu_l \ ,
\label{mu}
\end{eqnarray}
where ${Q_B}_i$ is the baryon number ($1/3$ for quarks and $0$ for
leptons) and ${Q_l}_i$ the lepton number ($0$ for quarks and $1$ for
leptons) of each particle. Note that the lepton chemical potential is
non-zero only when the lepton fraction is fixed. If strangeness were conserved, like in the case of the initial stages of a heavy ion collision, a strange chemical potential would have to be introduced.

Due to energy level quantization in directions perpendicular to the magnetic field, the thermodynamic quantities for charged particles are summed over the Landau levels $\nu$, which for zero temperature have a maximum value equal to $\nu_{max}=\lfloor(\mu_i^2-m_i^2)/(2|{Q_e}_i|eB)\rfloor$. Note that the degeneracy of the zeroth Landau level is always half of the usual value for spin $1/2$ particles. For non-charged particles, such as neutrinos, the pressure and energy density expressions take the usual form \cite{Chodos:1974je}.

For the star calculations, we require that the matter is charge neutral and in chemical equilibrium. The first condition can be written as
\begin{eqnarray}
\sum_i {Q_e}_i \rho_i=0 \ ,
\end{eqnarray}
where the density of each charged particle for finite and zero temperature cases is
\begin{eqnarray}
&& \rho_i=\sum_\nu \frac{\gamma_i(\nu)}{2 \pi^2}|{Q_e}_i|eB \int ({f_+}_i-{f_-}_i) dk \ ,
\\
&& \rho_i(T=0)=\sum_\nu \frac{\gamma_i(\nu)}{2 \pi^2}|{Q_e}_i|eB \bar{k}_{Fi} \ ,
\end{eqnarray}
and the total baryon density is simply $\sum_i {Q_B}_i \rho_i$.  The
second condition, corresponding to chemical equilibrium, can be
written by replacing the baryon number, electric charge, and lepton number for each particle in Eq.~(\ref{mu})
\begin{eqnarray}
\label{ddd}
&&\mu_u=\frac 1 3 \mu_B+ \frac 2 3 \mu_q+ \frac 2 3\mu_l \ , \ \ \ \ \ \mu_d=\frac 1 3\mu_B- \frac 1 3\mu_q- \frac 1 3\mu_l \ ,\nonumber
\\
&&\mu_s=\frac 1 3\mu_B- \frac 1 3 \mu_q- \frac 1 3\mu_l \ , \ \ \ \ \ \ \mu_e=-\mu_q \  , \nonumber
\\
&&\mu_\mu=-\mu_q \ , \ \ \ \ \  \mu_{\nu_e}=\mu_l \ , \ \ \ \ \ \mu_{\nu_\mu}=\mu_l \ .
\end{eqnarray}
It is easy to see from Eq.~(\ref{ddd}) that
$\mu_d=\mu_s=\mu_u+\mu_e-\mu_l$, $\mu_e=\mu_\mu$, and $\mu_{\nu_e}=\mu_{\nu_\mu}$. Note that the quark chemical potentials depend on the lepton chemical potential, even though they do no posses lepton number. This is a consequence of fixing the lepton fraction as explained in Ref.~\cite{Hempel:2009vp}.

Due to the inclusion of the magnetic field, the components of the energy-momentum tensor $T_{11}$, $T_{22}$ and $T_{33}$ are further modified. In the literature, components $T_{11}$ and $T_{22}$ are usually referred to as perpendicular pressures, and component $T_{33}$ as parallel pressure. In this work we follow this nomenclature. In fact, one can show that the difference of the parallel and perpendicular components of the pressure
is proportional to the magnetization. The magnetization can be calculated by taking the negative of the derivative of the grand potential with respect to the magnetic field, resulting for finite and zero temperature in
\begin{eqnarray}
M&=&-\frac{\partial \Omega}{\partial B}=\frac{P_{m_\|}}{B}-\sum_{i,\nu} \frac{\gamma_i(\nu)}{2 \pi^2}{Q_e}_i^2e^2B \nu\int \frac{1}{\sqrt{k_i^2+\bar{m}_i^2}} ({f_+}_i+{f_-}_i) dk \ ,
\end{eqnarray}
\begin{eqnarray}
M(T=0)=\frac{P_{m_\|}}{B}-\sum_{i,\nu} \frac{\gamma_i(\nu)}{2 \pi^2}{Q_e}_i^2e^2B \nu \ln{\left|\frac{\bar{k}_{Fi}+\mu_i}{\bar{m}_i}\right|} \ ,
\end{eqnarray}
so the perpendicular component of the matter contribution to the pressure becomes
\begin{eqnarray}
P_{m_\perp}=P_{m_\|}-MB \ .
\label{eq:paniso}
\end{eqnarray}

In this context we note that a fair amount of debate exists in the literature concerning the existence
or non-existence of pressure anisotropies in the matter contribution to the energy-momentum tensor.
The early works of Canuto and Chiu (see e.g. Refs.~\cite{Canuto:1969ct,Canuto:1969cs}) explicitly 
demonstrated that, in the presence of a background magnetic field, a Fermi-gas of spin-one-half particles
possesses a pressure anisotropy.  Recently, we have revisited this calculation, including the effects
of anomalous magnetic moment \cite{Strickland:2012vu}, explicitly proving that Eq.~(\ref{eq:paniso})
holds for both charged and uncharged particles, with and without anomalous magnetic.  This calculation
was performed, starting from the Lagrangian, solving the Dirac equation, and explicitly performing the
statistical averages necessary to obtain the bulk components of the energy-momentum tensor.  In other words, Eq.~(\ref{eq:paniso}) can be derived without any underlying assumptions from
classical electromagnetism.

The confusion in the literature stems from the final step in which one expresses the transverse pressure
in the form (\ref{eq:paniso}).  It was first speculated by Blandford and Hernquist \cite{1982JPhC...15.6233B} 
that due to the presence of a non-vanishing magnetization that one needed to additionally take into account 
the Lorentz force of the external magnetic field 
on the bound current densities, ${\bf J}_m = \nabla \times {\bf M}$.  Recently this 
argument has reappeared in the literature \cite{Potekhin:2011eb} where it was once again argued that
one should take into account the bound current densities and also the bound surface current density
${\bf K}_m = ({\bf M} \times \hat{\bf n}) \, \delta_{\rm surface}$, where $\delta_{\rm surface}$ represents a
one-dimensional delta function which restricts the current to the surface of the material and $\hat{\bf n}$ is
the local surface normal vector.  As a result, the authors would argue that, if one considers an infinitesimal volume
$dV$, that there is an additional Lorentz force $d{\bf F}_m = d{\bf j}_m \times {\bf B}_{\rm ext}$ where 
$d{\bf j}_m = ({\bf J}_m + {\bf K}_m) \, dV$.  However, in the context of a gas of ionized particles subject to an external
magnetic field, adding such a contribution would amount to double-counting as we will now explain.

When one treats the problem of magnetized matter as a
self-bound dielectric, the underlying magnetization emerges from the orbits of electrons which are bound to charge
centers (nuclei) by {\it internal forces}.  When an external magnetic field is applied, the dipole moments of 
these (already bound) charges are rotated causing the system to generate a coherent magnetic field.
As is familiar from textbooks on this subject, density variations can then cause there to be a residual
bound internal current (${\bf J}_m$) and a residual bound surface current (${\bf K}_m$).
Since the magnetic field was not what was  keeping them bound in circular orbits in the first place, in this
case it makes perfect sense that one should then take into account the effect of the Lorentz force
on the residual currents induced by the orbiting electrons (since this interaction was not already accounted for).  
However, this is not the case when one has an ionized Fermi gas subject to an external magnetic field.  In this case, 
it is the external magnetic field itself which is keeping the particles in their quantized orbits and to then claim that one should 
once again include the effect of the {\it external} Lorentz force on these particles would be double-counting.

Along these lines we point out that for the case of self-bound dielectric matter with uniform density perpendicular to a constant magnetic field, there is only the surface magnetization current ${\bf K}_m$ on the edge of the planes perpendicular to the magnetic field.  
In this case one receives a contribution of $P_\perp = M B \, \delta_{\rm surface}$ to the pressure, which acts to restore isotropy in an infinitesimal shell on the surface.  
However, this does not change the fact that in the bulk of the material the pressures are anisotropic.  Since one has to establish mechanical equilibrium at each point in the volume, one must take into account pressure anisotropies even in the case of self-bound dielectric matter.

Having established the existence of pressure anisotropy in the matter contribution via Eq.~(\ref{eq:paniso}), 
we now turn to the pure-field contributions to the energy-momentum tensor (not related to the magnetization of the system).  Due to the presence of the external 
magnetic field, there are contributions from the electromagnetic field tensor (see Eq.~(\ref{lag})) to the pressures 
and energy density \cite{kapusta,1982JPhC...15.6233B}.  Due to the breaking of rotational symmetry
by the magnetic field, these contributions are again different in the parallel 
and perpendicular directions. The resulting pressures and energy density are
\begin{eqnarray}
\epsilon&=&\epsilon_{m}+\frac{B^2}{8\pi} \ , \ \ \ \ \ P_{\perp}=P_{m_\perp}+\frac{B^2}{8\pi} \ , \ \ \ \ \ P_{\|}=P_{m_\|}-\frac{B^2}{8\pi} \ ,
\end{eqnarray}
where the $4\pi$ comes from the choice of Gaussian natural units.

\begin{figure}[t]
\centering
\includegraphics[width=0.55\textwidth]{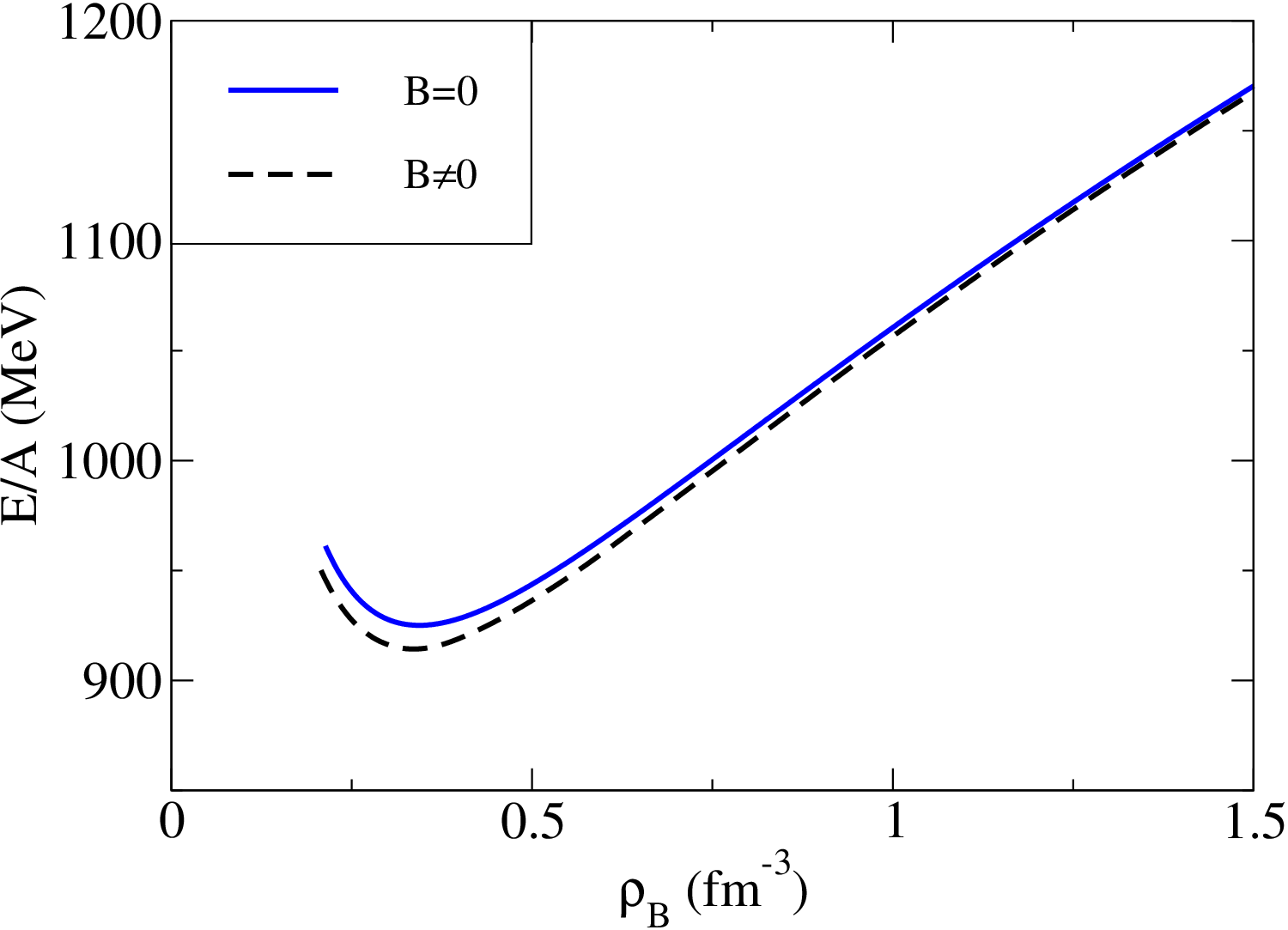}
\caption{(Color online) Binding energy versus baryon density for charge neutral matter at zero temperature shown without and with (constant) magnetic field effects. For the magnetized case $B=4.3\times10^{18}$ G.}
\label{bind}
\end{figure}

The temperature is not expected to be constant in the interior of
compact stars. This can be easily taken into account by
computing quantities at fixed entropy per particle. Numerical simulations
\cite{Burrows:1986me, Pons:2001ar,pons/pons,Stein:2005nt} show that this ratio can reach values 
1 or 2, depending on the stage of evolution. In this case,
the temperature increases naturally from a smaller value on the surface to a higher value in the center of the star. The entropy per particle (baryon) can be calculated through the thermodynamical expression
\begin{eqnarray}
\frac{S}{A}= \frac{s}{\rho_B}=\frac{\epsilon+P-\mu_B\rho_B}{T \rho_B} \ .
\end{eqnarray}

\begin{figure}[t]
\centering
\includegraphics[width=0.55\textwidth]{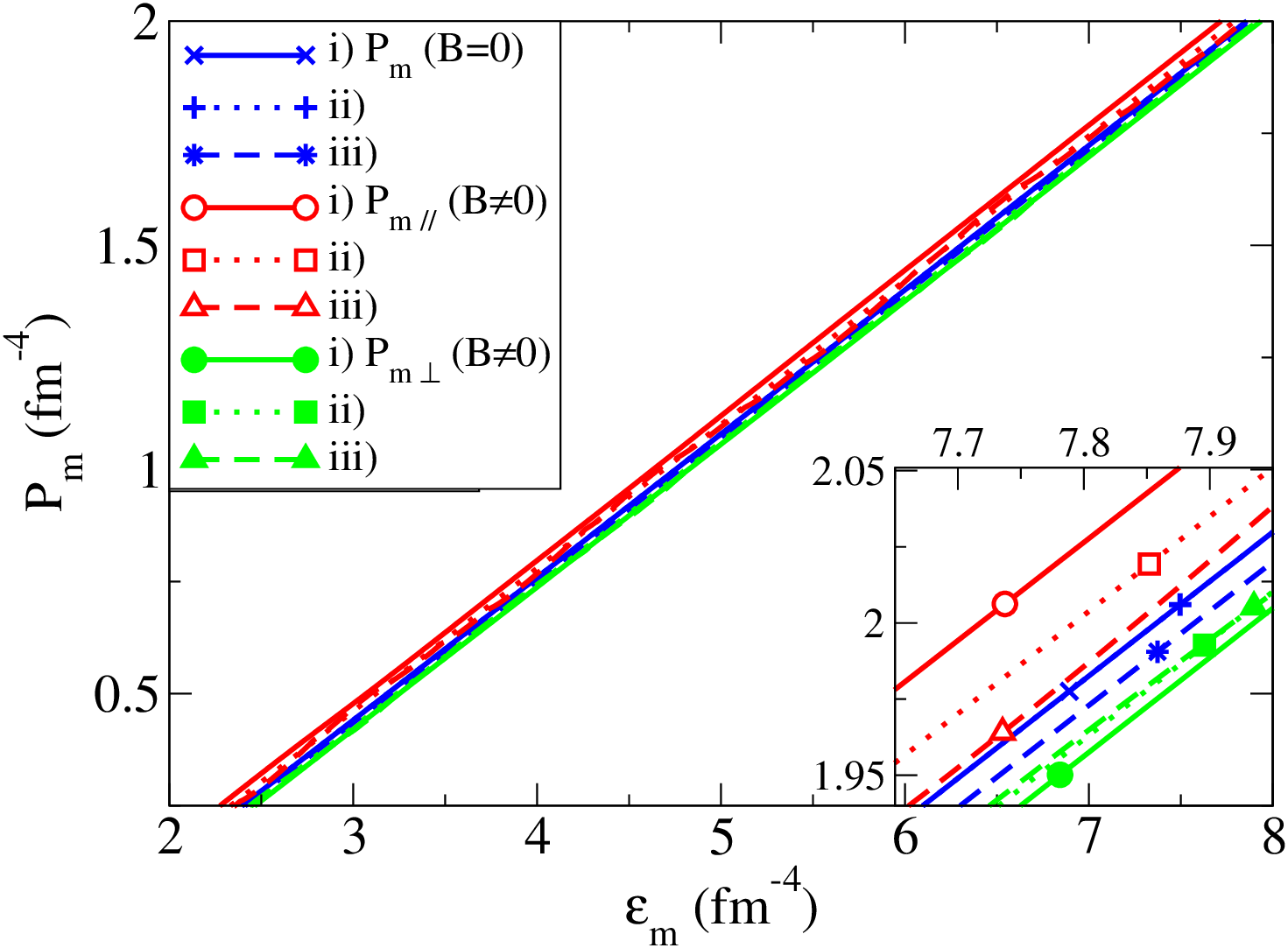}
\caption{(Color online) Different matter contributions to the pressure versus matter contribution to the energy density at three snapshots of the star evolution shown without and with (constant) magnetic field effects. For the magnetized case $B=4.3\times10^{18}$ G.}
\label{eos}
\end{figure}

Immediately after the supernova explosion, the neutrinos are still trapped \cite{Gudmundsson:1980, Keil:1995hw}, which can be modeled using a fixed lepton fraction defined as 
\begin{eqnarray}
Y_l=\frac{\sum_i {Q_l}_i \rho_i}{\rho_B} \ ,
\end{eqnarray}
which according to simulations \cite{Burrows:1986me} is
$Y_l\simeq0.4$. Here we assume $Y_l=0.4$ throughout. 
Note that more sophisticated approaches have realistic profiles
  for temperature and lepton fraction \cite{Sanjay:1998, Pons:2001}. 
In this work we do not attempt to make use of them since our aim is
only to investigate magnetic field effects on the star evolution. Note that in this work we assume that both kinds of neutrinos, electron and muon,  are initially trapped and concurrently emitted. If, instead, we only include electrons and electron neutrinos in the lepton fraction, we find that the results are qualitatively the same and quantitatively similar. The small quantitative difference stems from the fact that there are not many muons present in the star.

\section{Results and discussions}
\label{sect:results}

We begin our analysis with the binding energy for matter without
leptons. The results at zero temperature are shown in
Fig.~\ref{bind}. In this work, as previously stated, we use a bag
constant of ${\mathcal B}=(154\ \rm{MeV})^4$, which reproduces a minimum in the
binding energy ($B/A=925.9$
MeV) slightly below the one for the iron. We point out that the inclusion of the magnetic field makes matter even more stable, giving a binding energy that is even lower ($B/A=914.30$ MeV). Such a behavior was already pointed out in Refs.~\cite{Felipe:2008cm,Wen:2012jw}. 

We now proceed with our analysis of the equation of state of the matter only (the pure magnetic field contributions of $\pm B^2/8\pi$ are not included at first). The parallel and perpendicular pressures of matter are plotted in Fig.~\ref{eos} as a function of energy density of matter for three different cases:
\begin{enumerate}[i)]
\item $s/\rho_B=1$, $Y_l=0.4$ \ ,
\item $s/\rho_B=2$, $\mu_{\nu_l}=0$ \ ,
\item $s/\rho_B=0$, $\mu_{\nu_l}=0$ \ .
\end{enumerate}
These correspond to three snapshots of the time evolution of a quark star in
its first minutes of life. Such an analysis is important, as it has
been shown in calculations/simulations that quark matter can already
be formed during supernova explosions
\cite{Sagert:2008ka,Fischer:2011zj}. At first, the star is relatively
warm (represented by fixed entropy per particle) and has a large number of
trapped neutrinos (represented by fixed lepton fraction). As the trapped 
neutrinos diffuse, they heat up the star due to Joule heating \cite{Prakash:1996xs,Burrows:1986me}. Finally, the star can be considered cold. In reality, the situation may be more complicated because the entropy profile of the star is not homogeneous. In Ref.~\cite{Burrows:1986me}, for example, it was found that the central entropy initially increased as a function of time, however, the integrated total entropy per baryon decreased with time.

\begin{figure}[t]
 \centering
\includegraphics[width=0.55\textwidth]{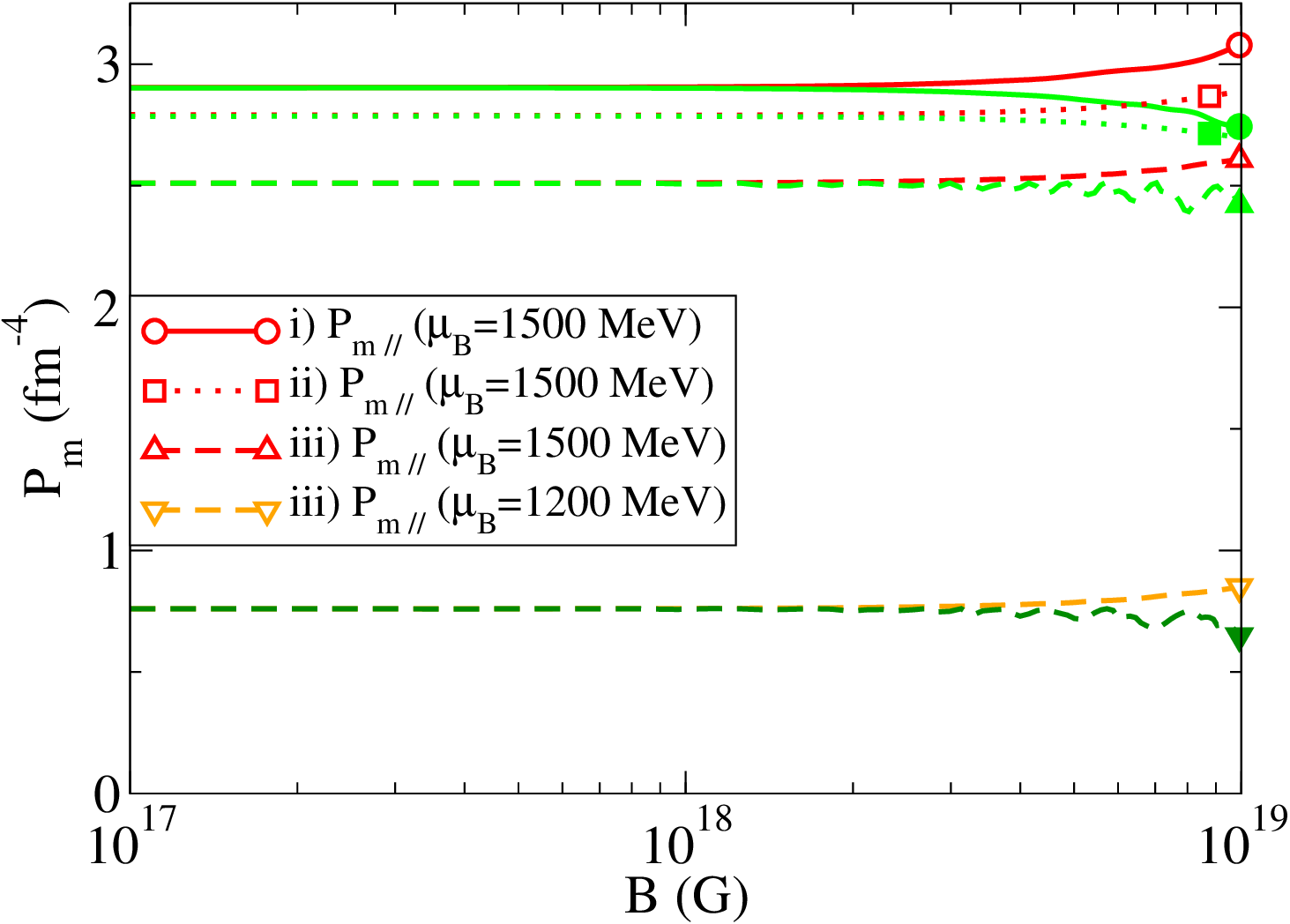}
 \caption{(Color online) Parallel (open symbols on red/orange lines) and perpendicular
   (respective full symbols on green/ dark green lines) pressures of matter versus
   magnetic field representing three snapshots of the star
   evolution.}
 \label{b2}
\end{figure}

Notice in Fig.~\ref{eos} that (constant)
magnetic field effects, seen by the difference between the parallel
and perpendicular pressures, are more pronounced in the beginning of
the star evolution (i), when there is a larger number of negatively
charged leptons and up quarks.  Other effects, such as the Haas-van
Alphen oscillations \cite{a1,a2,a3,a4} can only be seen at the last
stage of evolution (iii), since such oscillations are smoothed out by thermal
effects. These oscillations can be better seen in Fig.~\ref{b2} where
the parallel and perpendicular pressures of matter are plotted as a
function of magnetic field for different chemical potentials
representing the center of the star ($\mu_B=1500$ MeV) and an outer
region ($\mu_B=1200$ MeV). Our results for $T=0$ can be compared with the 
ones presented in Fig.~1 of Ref.~\cite{Paulucci:2010uj}. The models are not the
  same, but the same physics is obtained for $\mu_B=1500$ MeV (which
  corresponds to $\mu_q=500$ MeV in Ref. \cite{Paulucci:2010uj}).
 At these high densities, the matter contribution to the perpendicular
  pressure only deviates significantly from the parallel one when the magnetic
  fields are as high as $3 \times 10^{19}$ G.
  
 \begin{figure}[t]
 \centering
\includegraphics[width=0.55\textwidth]{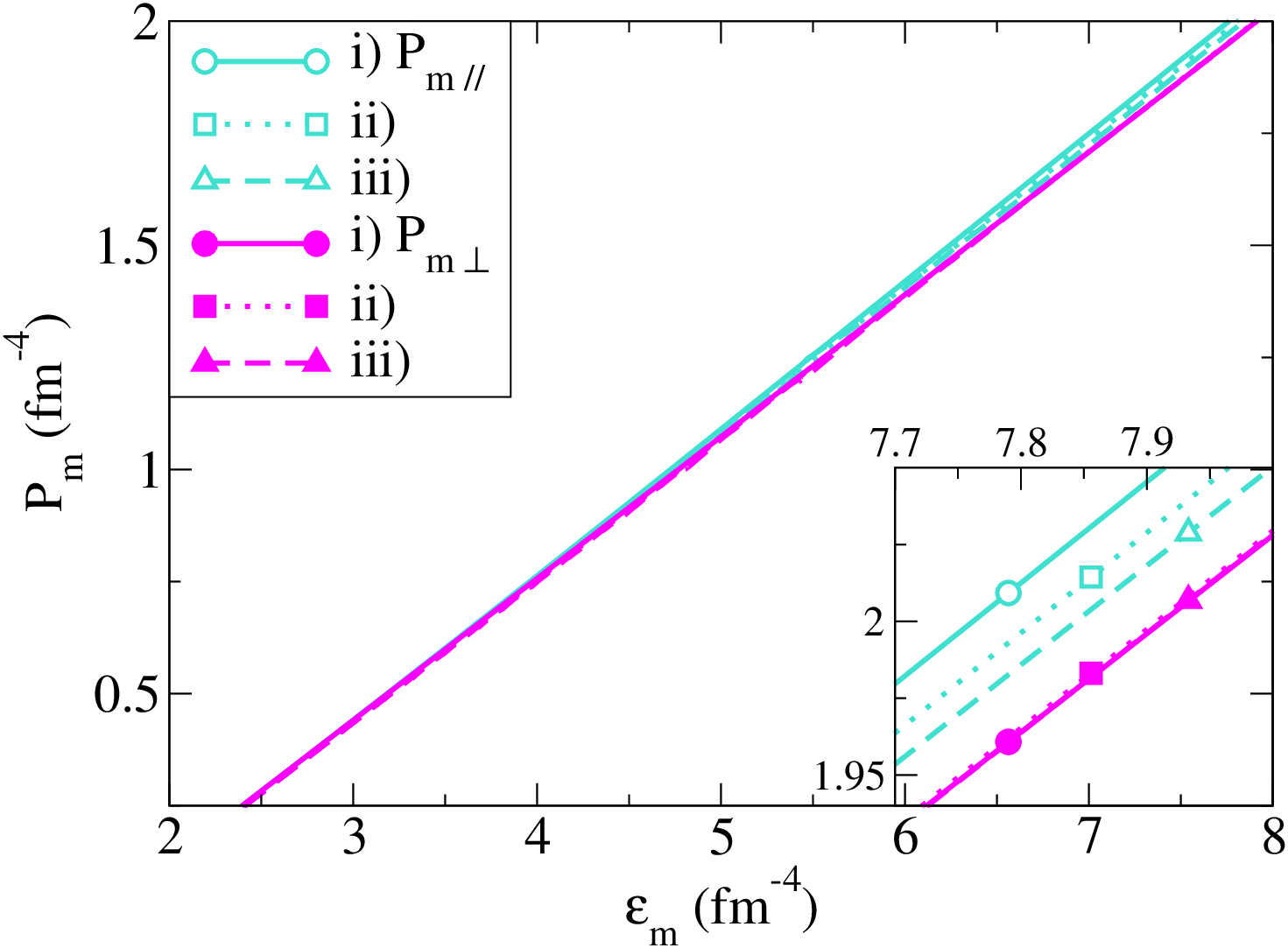}
 \caption{(Color online)  Different matter contributions to the pressure versus matter contribution to the energy density at three snapshots of the star evolution shown for variable (chemical potential dependent) magnetic field with $B_c=4.3\times10^{18}$ G.}
 \label{eos2}
\end{figure}

The magnetic field configuration considered in Figs.~\ref{bind} and \ref{eos} was constant in the
$z$ direction with $B=4.3\times10^{18}$ G.  One expects a non-constant magnetic
field whose magnitude decreases as one goes out from the center towards the edge of the star, in accordance with full general relativity calculations \cite{Bocquet:1995je,Cardall:2000bs} .  To allow for this, we consider a 
magnetic field with a variable strength $B^*(\mu_B)$ that increases with chemical potential (density). It
ranges from a lower value set to $B_{surf}=10^{15}$ G at small baryon
chemical potentials (densities) to a higher value set to
$B_c=4.3\times10^{18}$ G at high baryon chemical potentials
(densities) following the prescription from 
Refs.~\cite{Bandyopadhyay:1997kh,Mao:2001fq,Rabhi:2009ih,Dexheimer:2011pz}
\begin{equation}
B^*(\mu_B)=B_{surf}+B_c \left[1-e^{b\frac{(\mu_B-938)^a}{938}}\right] \ ,
\label{B}
\end{equation}
with $a=2.5$, $b=-4.08\times10^{-4}$ and $\mu_B$ given in MeV.  These constants are related to the rate of growth of the magnetic field with the baryon chemical potential. 
Changing these constants (within reason) does not qualitatively change the results (see e.g. Ref.~\cite{Rabhi:2009ih,Mukhopadhyay:2013yra} where
the authors studied the dependence of the EOS on the choice of these parameters).
The variable magnetic field prescription employed only  significantly modifies the behavior of the system at high densities and therefore does not change the minimum of the binding energy. As explained in \cite{Ferrer:2010wz}, stronger magnetic fields are necessary at larger densities in order to conserve the magnetic flux. For an alternative explanation of how the magnetic field strength increases toward the center of the star see Ref.~\cite{Eto:2012qd}.

The effects of considering a variable magnetic field can be seen in Fig~\ref{eos2}, where the EOS is shown. They only become visible at high densities. It is important to notice that the magnetic field in the center of the star never reaches $B_c$, but only $\sim 80 - 90\%\ B_c$. This happens because the stars become gravitationally unstable for higher central chemical potentials. 

\begin{figure}[t]
\centering
\includegraphics[width=0.55\textwidth]{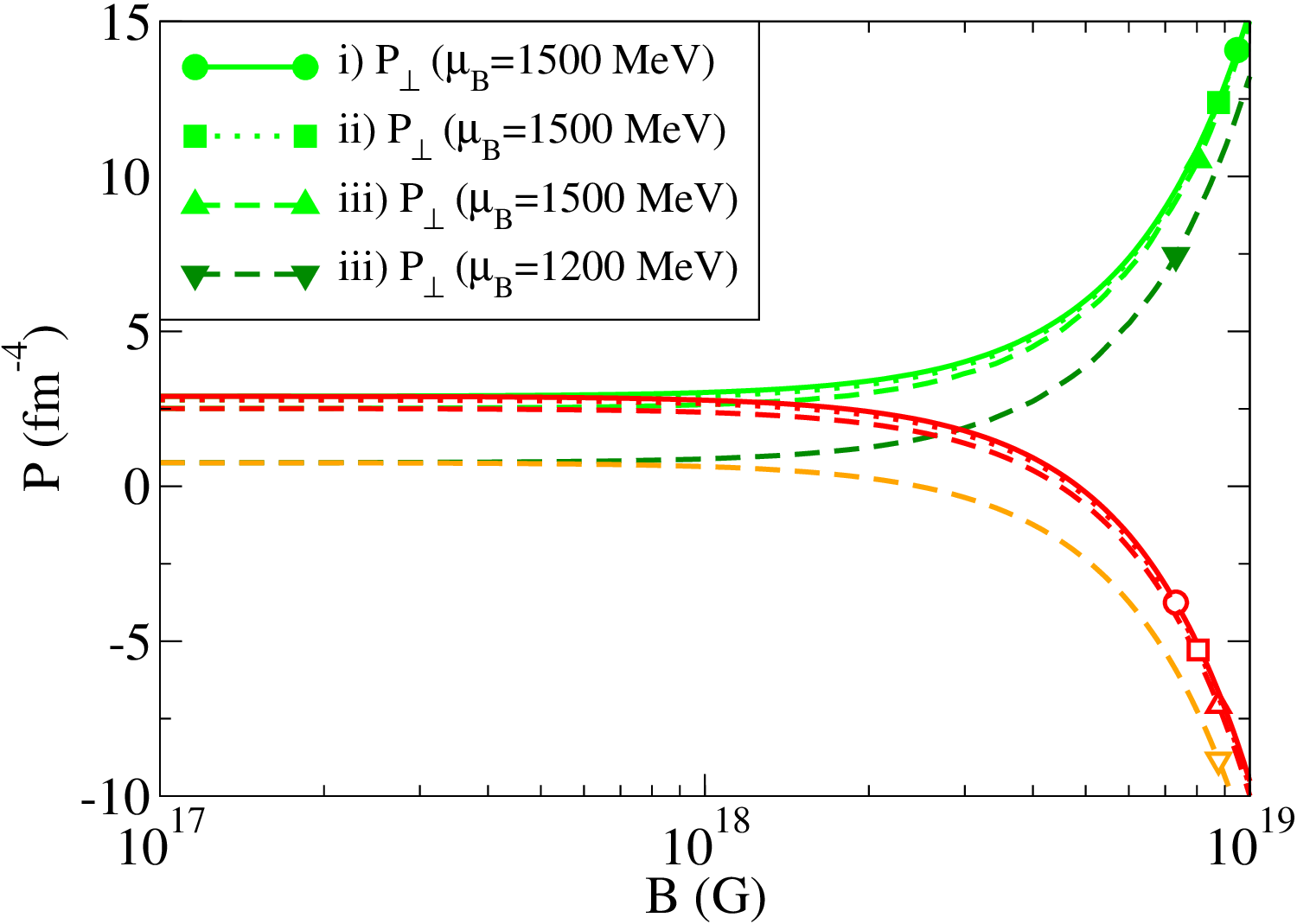}
\caption{(Color online) Parallel (open symbols on red/orange lines) and perpendicular (respective full symbols on green/dark green lines) pressures versus magnetic field representing three snapshots of the star evolution. Pure magnetic field contribution $B^2/8\pi$ included}
\label{b}
\end{figure}

We can ask ourselves if it would make sense to increase the
magnetic field further than we have already done. The answer is
probably not. Beyond $B_c^{\rm max}\equiv4.39\times10^{18}$ G the parallel pressure
(now including the field contribution) becomes negative in the
center of the star, even using the variable magnetic field
prescription. According to Ref.~\cite{Sinha:2010fm} this defines an
upper limit for the magnetic field, beyond which the system becomes
unstable. Using a constant magnetic field, the upper limit is
lower, as indicated in Fig.~\ref{b}. For a chemical potential of
$\mu_B=1500$ MeV the parallel pressure becomes negative for
$B=3.17\times10^{18}$ G but for $\mu_B=1200$ MeV this happens for a
lower magnetic field of $B=1.77\times10^{18}$ G. Those numbers were
obtained for stage (iii) of the star evolution. In the earlier stages,
the parallel pressure becomes negative for slightly higher magnetic
fields, $B=3.37\times10^{18}$ G for (i) and (ii) at $\mu_B=1500$ MeV. As this demonstrates, having
a variable magnetic field prescription improves the stability since it produces high 
magnetic fields only at high densities, where the longitudinal matter pressure is large enough to 
partially compensate for the negative pure magnetic longitudinal pressure. 

\begin{figure}[t]
\centering
\includegraphics[width=0.55\textwidth]{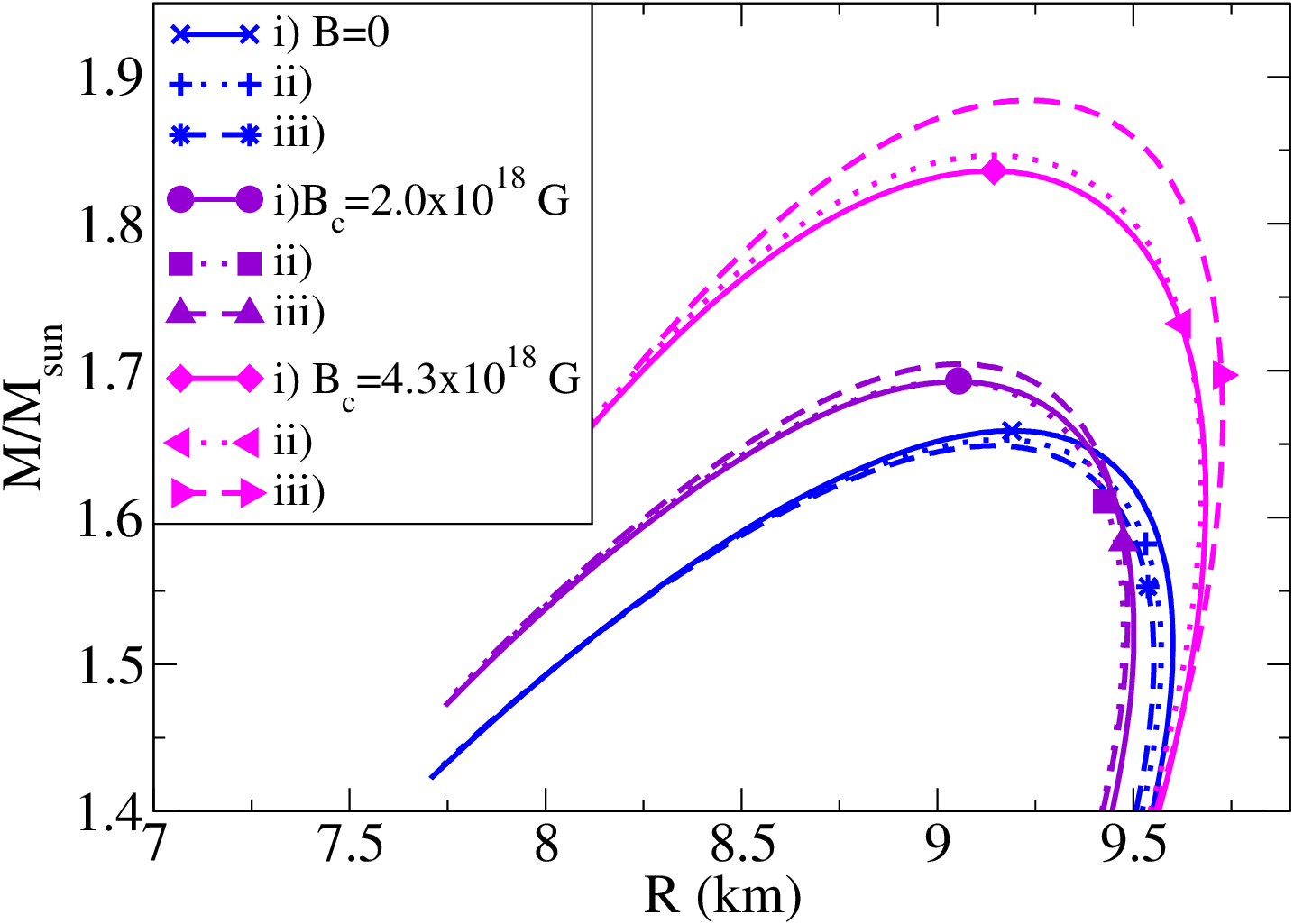}
\caption{(Color online) Mass-radius diagram for star families representing three snapshots of the star evolution shown without and with (variable) magnetic field. The symbols now represent the time evolution of the most massive stars considering fixed baryon number.  The lower part of the figure is not shown as it exhibits the usual behavior.}
\label{mass}
\end{figure}

Having obtained the EoS for the system, we use
this as input to the TOV equations \cite{Tolman:1939jz,Oppenheimer:1939ne} in order to
obtain an estimate of the effect of magnetic fields on the different stages of proto-quark star evolution.    
Note that we use the transverse pressure to solve the TOV equations.  This should be a reasonable approximation when 
$P_\perp \approx P_\parallel$; otherwise, one should also include the breaking of spherical 
symmetry in the solution of the gravitational metric.  We will consider
two different central magnetic fields of $B_c = 2.0 \times 10^{18}$ G and $B_c = 4.3\times10^{18}$ G,
the latter of which is close to $B_c^{\rm max}$ and whose reliability will be discussed later.
The results can be seen in the mass-radius diagram shown in  Fig.~\ref{mass}. Without magnetic field effects, the maximum mass that stars can have decreases with time from (i) to (ii) as the star deleptonizes and becomes more isospin symmetric (due to charge neutrality). Actually, the star has more up than down quarks at (i). From (ii) to (iii) the star's maximum mass decreases again as it becomes cold and loses thermal energy.
On the other hand, when the (variable) magnetic field effects are included, the maximum mass increases with time, from (i) to (ii) and from (ii) to (iii). The bulk of this result comes from the fact that thermal effects cancel out some of the magnetic field effects through changes in the particle population. When the (variable) magnetic field effects are included, the difference in the possible star masses between different snapshots is also larger and all stars are more massive.

The analysis above works only for stars that are in binary systems, and therefore can change baryon number with time. 
So far we have not taken into account baryon number conservation.  Our results for the equation-of-state itself and
pressure anisotropy are general but, in order to analyze the dynamics of an isolated star, we need to constrain the star's baryon number to be the one it had at the first moment of evolution (i). In this way, we can consistently determine its properties such as mass and radius at different points in its evolution.  As can be seen in Table \ref{table} the maximum masses of isolated stars in all cases now decrease with time. This result agrees with similar hadronic calculations \cite{Dexheimer:2008ax}. It is interesting to note that within this analysis, the magnetic field in the center of isolated stars from stages (ii) and (iii) only reaches about up to $\sim 40 - 50\%\ B_c$. The data from Table~\ref{table} with fixed baryon number is also represented in Fig.~\ref{mass}, where the symbols represent the evolution of the most massive stars allowed (with baryon number determined by stage (i)). 

\begin{figure}[t]
\centering
\includegraphics[width=0.55\textwidth]{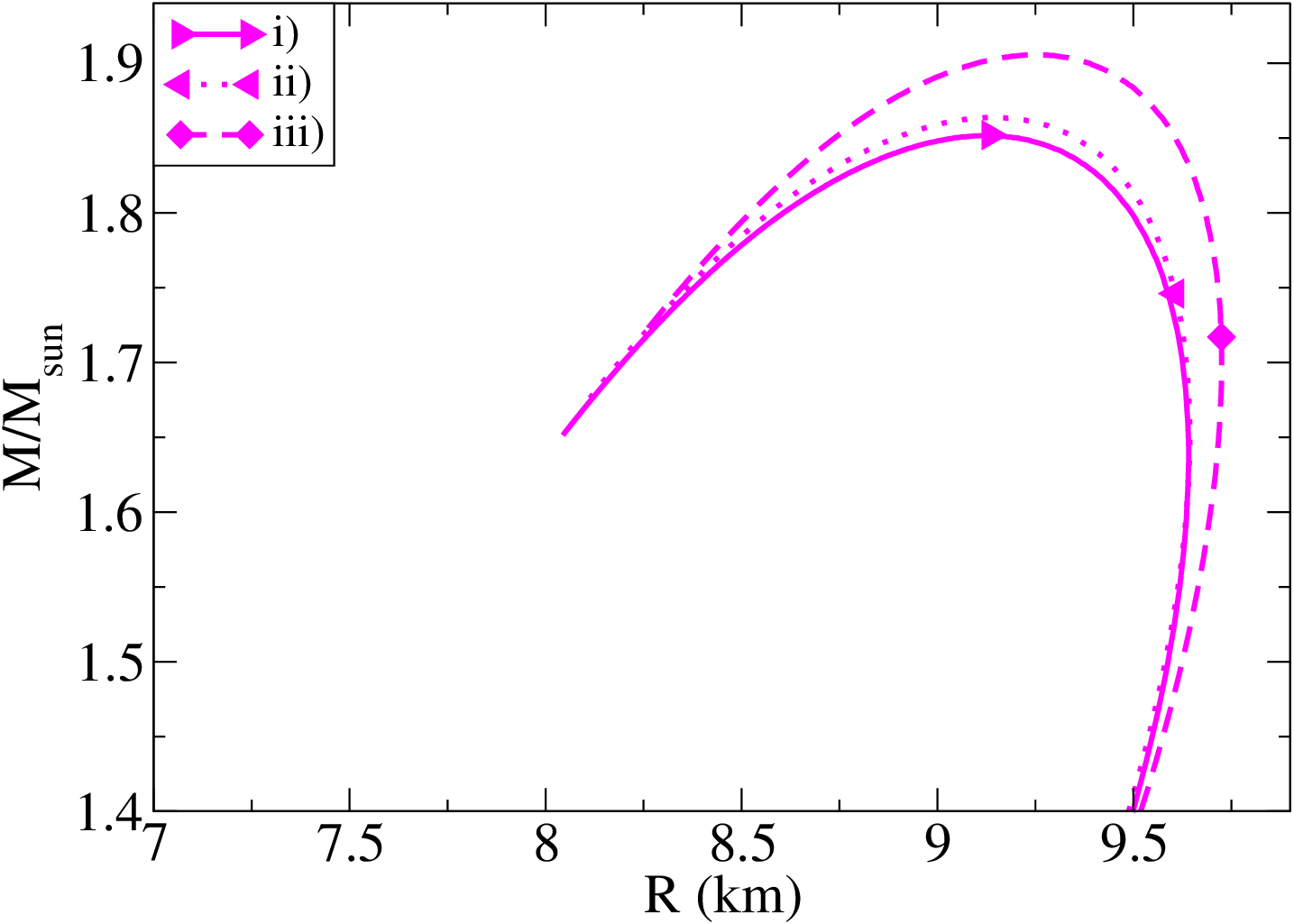}
\caption{(Color online) Mass-radius diagram for star families representing three snapshots of the star evolution shown with variable magnetic field using  $B_c = 4.3 \times 10^{18}$ G and $B_{surf} = 4.3 \times 10^{17}$ G. The symbols again represent the time evolution of the most massive stars considering fixed baryon number.  The lower part of the figure is not shown as it exhibits the usual behavior.}
\label{mass2}
\end{figure}

In addition, we would like to point out that our results are not qualitatively dependent on amount by which the magnetic field increases from the surface to the center of the star. Fig.~\ref{mass2} shows that increasing the magnetic field by less than one order of magnitude instead of three orders of magnitude as done in Fig.~\ref{mass}, the conclusion remains the same: the maximum masses of isolated stars in all cases decrease with time.  In order to demonstrate this, in Fig.~\ref{mass2} we have changed $B_{surf}$ from $10^{15}$ G (the maximum value observed on the surface of stars) to $4.3 \times 10^{17}$ G, a value closer to the ones from full general relativity calculations from Refs.~\cite{Bocquet:1995je,Cardall:2000bs}, for comparison. Note that the ratio between the magnetic field in the center of the star and on the surface is not simply $B_c/B_{surf}$, but less, since the central magnetic field in the star never reaches $B_c$.

It is important to note that we assume that the magnetic field is constant in time in the analysis of different snapshots of the quark star evolution. Such an assumption is reasonable considering that the timescale for the magnetic field decays is of the order of $10^5-10^7$ years \cite{Harding:2006qn,Aguilera:2007xk,Pons:2007vf,Xie:2011bd}. This should be contrasted with the timescale for the cooling of proto-neutron stars which is on the order of minutes.

\begin{table}
\caption{ Maximum masses (without and with fixing baryon number), corresponding baryon number and corresponding radii for different snapshots of the star's evolution without and with the inclusion of a spatially varying magnetic field}.  For the case with magnetic field we used $B_c=2.0\times10^{18}$ and $B_c=4.3\times10^{18}$ G. \label{table}
%\begin{indented}
%\item[]
\renewcommand*{\arraystretch}{1}
{\begin{tabular*}{\textwidth}{c|ccc|ccc|ccc}
\br
~&\multicolumn{3}{c}{$B=0$}&\multicolumn{3}{c}{$B_c=2.0\times10^{18}$}&\multicolumn{3}{c}{$B_c=4.3\times10^{18}$}\\
{stages}&i&ii&iii&i&ii&iii&i&ii&iii\\
\mr
$\rm{M}_{\rm{max}}(\rm{M}_\odot)$&{\small 1.66}&{\small 1.65}&{\small 1.65}&{\small 1.69}&{\small 1.69}&{\small 1.71}&{\small 1.84}&{\small 1.85}&{\small 1.88}\\
{\small A($\times10^{58}$)} &{\small 0.222}&{\small 0.234}&{\small 0.238}&{\small 0.225}&{\small 0.238}&{\small 0.245}&{\small 0.244}&{\small 0.260}&{\small 0.270}\\
{\small $\rm{M}_{\rm{max}}(\rm{M}_\odot)$ fixed A}&{\small 1.66}&{\small 1.58}&{\small 1.55}&{\small 1.69}&{\small 1.61}&{\small 1.58}&{\small 1.84}&{\small 1.73}&{\small 1.70}\\
A($\times10^{58}$) fixed A &{\small 0.222}&{\small 0.222}&{\small 0.222}&{\small 0.225}&{\small 0.225}&{\small 0.225}&{\small 0.244}&{\small 0.244}&{\small 0.244}\\
%R(km) fixed A &9.19&9.53&9.54&9.06&9.43&9.47&9.14&9.62&9.73\\
\br
\end{tabular*}}
%\end{indented}
\end{table}

As a final remark, another option would be to make use of a non-static magnetic field profile that changes in order to conserve the magnetic flux. To achieve this, one would need to iteratively solve the Tolman-Oppenheimer-
Volkoff Equations, since the star radius depends on the magnetic field self-consistently. Although
this is a very interesting idea, it is beyond the scope of the
current work, which was to analyze the effect of strong magnetic fields on snapshots of the evolution of
an isolated quark star for fixed magnetic field. That being said, based on the results already contained
in Table I, for the largest magnetic field considered between stages (i) and (iii) the
surface area of the star changes by approximately 13\%. Assuming conservation of magnetic flux
would imply that the magnetic field would be reduced by approximately 12\% between stages (i) and
(iii). As a result, in the case of fixed baryon number we would expect this to enhance the effect we
found, since the maximum mass decreases with decreasing magnetic field. Based on our results, we
would expect the size of the effect to be on the order of a 1.5\% reduction in the maximum mass of the
star for the case of fixed baryon number and the maximum magnetic field studied. In the case when the
baryon number is allowed to change, this effect could further reduce the already small dependence of
the star maximum mass on the magnetic field.

\section{Conclusions and Outlook}
\label{sect:conclusions}

In this paper we have examined the effect of strong magnetic fields on proto-quark stars 
using a simple quark model.  Three situations were investigated: (i) fixed entropy per baryon 
$s/\rho_B=1$ with trapped neutrinos, (ii) $s/\rho_B=2$ without neutrinos and (iii) a zero
temperature system with no neutrinos.  The pressure anisotropy inherent
to Fermi gases in external magnetic fields was obtained both for fixed magnetic 
fields and for a scenario where the magnetic field varies with the baryon chemical 
potential.  The conditions for chemical equilibrium between the quarks, leptons, and
neutrinos were enforced.  In the case of isolated proto-quark stars we incorporated the
effect of baryon number conservation during their evolution.

We demonstrated that strong magnetic fields modify quark star masses. 
However, the evolution of isolated stars needs to be constrained
by fixed baryon number (from the first stage (i)), which lowers the possible star masses for each equation of state.  For our final mass
vs radius results we studied two different central magnetic fields: $B_c = 2.0\times 10^{18}$ G
and $B_c = 4.3\times 10^{18}$ G.  In both cases the maximum magnetic field 
generated in the star's interior at stage (i) of the evolution is on the order of $B_{\rm max} \sim 0.8 B_c$.  For
the lower value of the magnetic field considered, this places us well below $B_c^{\rm max}$
where the longitudinal pressure of the system becomes negative.  In addition, the level
of pressure anisotropy in this case is relatively small with $P_\parallel/P_\perp
\simeq 0.85$, giving us some confidence in the use of isotropic TOV equations.  For the larger value of the magnetic field studied,
we find that the longitudinal pressure is still positive; however, the level of pressure anisotropy 
is quite large with $P_\parallel/P_\perp \simeq 0.4$.  In this second case, it is highly questionable whether isotropic TOV is sufficient
to draw firm numerical conclusions; however, as we have demonstrated, the qualitative behavior 
and pattern of evolution between the different stages observed with this larger magnitude of the magnetic field are similar
to those obtained with the smaller magnitude magnetic field.  We have also demonstrated that the qualitative behavior of our results is also not dependent on the amount that the magnetic field increases throughout the star.

Additionally, our results show clearly that the MIT bag model for a ${\mathcal B}^{1/4}=154$
MeV obtained from an investigation of the adequate stability window
\cite{us} cannot reproduce the very massive neutron star recently
detected \cite{demorest}, not even if very intense magnetic fields are
considered. This is perhaps not surprising since most models which are able to
describe large mass stars rely on the existence of additional repulsive (vector) interactions \cite{Klahn:2006iw,Bonanno:2011ch}.
That being said, it is interesting to consider the
effect of a magnetic field in the somewhat simple bag model in order to establish a 
baseline for other models in a manner which does not introduce additional physical
effects such as, e.g. phase transitions whose nature might depend on the background magnetic fields, etc.

Finally, we note that although the results presented imply that quark stars with 
$B_c = 2.0 \times 10^{18}$ G are only mildly anisotropic, larger central magnetic fields
cause the pressures to become increasingly anisotropic.  Therefore, at such values of the magnetic
field one should solve Einstein's equations in an axisymmetric metric which is determined
self-consistently from the axisymmetric energy-momentum tensor for the star.  
Numerical solution for the axisymmetric case is needed and we are currently working 
towards this goal. Other issues to investigate in the future are the inclusion
of the effect of the anomalous magnetic moment and color
superconductivity. The later becomes very interesting in the
presence of strong magnetic fields \cite{Ferrer:2005vd,Ferrer:2006vw,Manuel:2006gu}.

\ack

V.D. and D.P.M acknowledge support from CNPq/Brazil.
M.S. was supported by NSF grant No.~PHY-1068765 and the Helmholtz International Center for FAIR LOEWE program.

\end{document}